\documentclass[aps,prl,twocolumn,superscriptaddress,amssymb,amsmath,nofootinbib]{revtex4-1}

\usepackage{amsmath,setspace,subfigure,amsfonts,latexsym}
\usepackage{amssymb}
\usepackage{color}
\usepackage{epsfig}
\usepackage{graphics}
\usepackage{sidecap}
\usepackage{slashed}
\definecolor{White}{rgb}{1,1,1}
\definecolor{Red}{rgb}{1,0.1,0}
\definecolor{LightYellow}{rgb}{1,1,.875}
\definecolor{SteelBlue}{rgb}{.273,.508,.703}
\definecolor{navy}{rgb}{0,0,.5}
\definecolor{LightCyan}{rgb}{.875,1,1}
\definecolor{DarkRed}{rgb}{.543,0,0}
\definecolor{HotPink}{rgb}{1,.41,.70}
\definecolor{ForestGreen}{rgb}{.13,.54,.13}
\definecolor{OliveDrab}{rgb}{.42,.55,.14}
\definecolor{MediumBlue}{rgb}{0,0,.80}
\definecolor{RoyalBlue}{rgb}{.25,.41,.88}
\definecolor{DeepSkyBlue}{rgb}{0,.746,1}
\definecolor{Brown}{rgb}{0.545,0.271,0.074}

\def\bea{\begin{eqnarray}}
\def\eea{\end{eqnarray}}
\def\bec{\begin{center}}
\def\ec{\end{center}}

\def\beq{\begin{equation}}
\def\eeq{\end{equation}}

\def\f{\frac}
\newcommand\lsim{\mathrel{\rlap{\lower4pt\hbox{\hskip1pt$\sim$}}
    \raise1pt\hbox{$<$}}}
\newcommand\gsim{\mathrel{\rlap{\lower4pt\hbox{\hskip1pt$\sim$}}
    \raise1pt\hbox{$>$}}}
\def\bea{\begin{eqnarray}}
\def\eea{\end{eqnarray}}
\def\ba{\begin{array}}
\def\ea{\end{array}}
\def\bc{\begin{center}}
\def\ec{\end{center}}
\def\nn{\nonumber}

\def\f{\frac}

\def\l{\lambda}

\def\f#1#2{\frac{#1}{#2}}

\begin{document}

\title{\Large Unstable Particles near Threshold}

\author{Dongjin Chway}
\email{djchway@gmail.com}
\affiliation{Department of Physics and Astronomy
and Center for Theoretical Physics, Seoul National University, Seoul 151-747, Korea}
\author{Tae Hyun Jung}
\email{thjung0720@gmail.com}
\affiliation{Department of Physics and Astronomy
and Center for Theoretical Physics, Seoul National University, Seoul 151-747, Korea}
\author{Hyung Do Kim}
\email{hdkim@phya.snu.ac.kr}
\affiliation{Department of Physics and Astronomy
and Center for Theoretical Physics, Seoul National University, Seoul 151-747, Korea}
\affiliation{Institute for Advanced Study, Princeton, NJ08540, USA}

\begin{abstract}
We explore physics of unstable particles when mother particle mass is around the sum of its daughter particle masses. In this case, the conventional wave function renormalization factor is ill-defined. 
We propose a simple resolution of the threshold singularity problem which still allows the use of narrow width approximation by defining branching ratio in terms of spectral density. The resonance peak and shape is different for different decay channels and no single decay width can be assigned to the unstable particles.
Non-exponential decay happens in all time scales.

\end{abstract}

\maketitle

\noindent
{\bf Introduction}
The narrow width approximation (NWA) has played an important role in studying unstable particles. Unstable particle states can not be asymptotic states of a scattering amplitude in order to keep unitarity and causality{\cite{Veltman:1963th}. Nevertheless, NWA allows similar treatments of unstable particles by factorizing full scattering cross sections of stable states into production and decay parts. In most practical situations, heavy off-shell calculations are immensely simplified with NWA.

When NWA is used for Standard Model calculations with realistic parameters, it is enough to take conventional wave function renormalization factor, $Z$ for unstable particles, whose inverse is defined by a real part of $G^{-1}$ differentiated by momentum square at physical mass square. Conventional choice of the physical mass is a zero of ${\rm Re}(G^{-1})$. They work when all the dressed propagators \cite{Dyson:1949bp} are well approximated by Breit-Wigner(BW) distribution \cite{Breit:1936zzb}.

However, this $Z$ is ill-defined in some examples beyond the Standard Model. In many cases, a self energy included in a dressed propagator is proportional to $\bar{\beta} \equiv \sqrt{(1-\f{(m_{a}+m_{b})^{2}}{p^{2}})(1-\f{(m_{a}-m_{b})^{2}}{p^{2}})}$ where $m_{a, b}$ are masses of particles propagating in the loop. 
This is because the phase space volume of decay is proportional to $\bar{\beta}$ and self energy and decay rate are closely related by the optical theorem. The classification of interactions providing the self energy with the same property is done later in this Letter. For simplicity, we discuss the problem with a scalar theory in the text.

For the self energy proportional to $\bar{\beta}$, the $Z^{-1}$ contains a term proportional to $1/\bar{\beta}$ which diverges as the physical mass approaches the threshold mass, $m_{a}+m_{b}$ from below. Taken faithfully, $Z\to 0$ means every production and decay of the unstable particle vanish and the particle becomes isolated from the theory no matter how strong the interaction is, which is nonsense. 

Solving the ill-defined $Z$ problem has been attempted mostly by using complex pole scheme\cite{Kniehl:1998fn,Kniehl:2002wn} which relates complex pole(s) on the second Riemann sheet to physical quantities: its real part to physical mass, imaginary part to decay rate, and residue to $Z$. After the complex pole was conjectured to have physical meaning \cite{Peierls:1955}, its gauge independence was shown in $Z$ boson in the Standard Model \cite{Stuart:1991xk} and scalars \cite{Grassi:2001bz}, and the scheme was employed to Higgs physics\cite{Passarino:2010qk}. 
However, using complex pole can be traced back to residue theorem for contour integral over lower half plane of the second Riemann sheet where, 
below a threshold, the analytically continued propagator $G_2$ defined in the second Riemann sheet deviates from the correct propagator $G$ which should have been used in exact calculation.

To understand the problem, it is important to know what really happens to the dressed propagator as the physical mass approaches a threshold mass. If the physical mass is near the threshold, the kinetic term can be represented by $p^{2}-m^{2} \propto \bar{\beta}^{2}$, while the self energy term is proportional to $\bar{\beta}$ around the peak. Thus the self energy is dominant near the peak and it changes the shape of the propagater to be totally different from BW distribution. We propose generalized narrow width approximation by defining branching ratio in terms of spectral density.

As the propagator $G$ changes, $\rho (p^2)$,
 the spectral density of K\"all\'en-Lehmann representation \cite{Kallen:1952zz,Lehmann:1954xi} 
also changes.
Unlike BW distribution which gives exponential decay with a rate of imaginary part of a complex pole, the survival probability, $P(t)\equiv \left|\int_0^\infty dS e^{-i \sqrt{S} t} \rho (S) \right|^2$ does not exponential decay. Deviation from exponential decay for very short or long time in quantum field theory is well known \cite{Schwinger:1960,Goldberger:1964}.
We show non-exponential decay pattern in middle range of time when most decay happens if they are at the threshold.

\noindent
{\bf Factorization}
Consider a full scattering cross section constructed with all external states by stable particles. For simplicity, assume that one Feynmann diagram (Fig. \ref{fullscattering}) dominantly determines the process
which contains an unstable particle state, $\phi$ that ends up being a final state $\lambda$. After inserting the identity, $\int dS \delta(S-p_{\phi}^{2}) \int d^{4}p_{\phi}\delta^{4}(p_{\phi}-p_{\lambda})\theta(p_{\phi}^{0})$ into the full scattering cross section, we obtain
\bea
&\sigma(&{\rm initial} \to1,2,\cdots,n, \lambda) \nn \\
&=&\int_{S_{min}}^{S_{max}} dS \, \sigma({\rm initial} \to m_1,\cdots,m_n,\sqrt{S}) \, \rho_{\lambda}(S)
\label{factorization}
\eea
where $S_{\rm max}$ ($S_{\rm min}$) is the biggest (smallest) possible invariant mass squared of $\lambda$ and
\bea
\rho_{\lambda}(S)\equiv |\langle 0 \mid \phi (0) \mid \lambda ; S \rangle|^{2}
\eea
which is a spectral density to a specific channel $\lambda$.
If $\lambda$ contains more than one particle, $\sqrt{S_{\rm max}}$ is $E_{\rm tot}-\sum_{i=1}^{n}m_i$ where $E_{tot}$ is total initial energy. When $n>1$, we can take $S_{min}=0$.

\begin{figure}[t]
\includegraphics[width=0.4\textwidth]{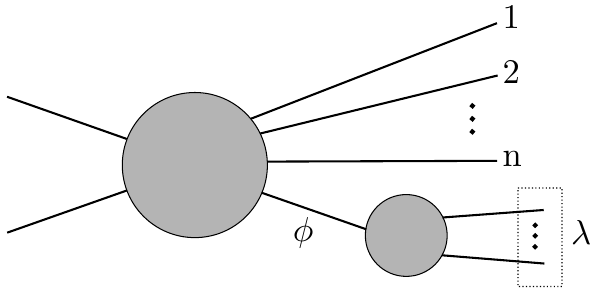}
\caption{Feynman diagram of initial $\to 1,2,\cdots,n,\lambda$.}
\label{fullscattering}
\end{figure}

If $\rho_{\lambda}(S)$ is delta function like with center $m_{\lambda}^{2}$ and width $m_\lambda \Gamma_{\lambda}$, and the production part, $\sigma({\rm initial} \to m_1,\cdots,m_n,\sqrt{S})$ does not change rapidly within the width, then we can approximate eq.(\ref{factorization}) by $\sigma({\rm initial} \to m_1,\cdots,m_n,m_{\lambda}) \int_{S_{\rm min}}^{S_{\rm max}} dS \rho_{\lambda}(S)$. For example, with momentum independent interaction, $\sigma({\rm initial} \to m_1, \sqrt{S})$
 is proportional to $\sqrt{(1-\f{(m_{1}+\sqrt{S})^{2}}{E_{\rm tot}^{2}})(1-\f{(m_{1}-\sqrt{S})^{2}}{E_{\rm tot}^{2}})}$. So for large total energy much higher than production threshold, the cross section has mild dependence on $S$. Relative error by taking it constant is of order $\f{m_{\lambda} \Gamma_{\lambda}}{E_{\rm tot}^{2}}$.

Production cross section is defined as the sum of full scattering cross sections of all different decay channels to detectable stable particles. Thus, it is a detection dependent concept and so is our wave function renormalization factor, $Z'$ which is defined as
\bea
Z'(E_{\rm tot};\Lambda)\equiv \sum_{\lambda \in \Lambda}\int_{S_{\rm min}}^{S_{\rm max}} dS \, \rho_{\lambda}(S)
\label{defZ}
\eea 
where $\Lambda$ is a set of relevent decay modes. If $E_{\rm tot}$ is taken to be infinity, $Z'(\infty;all)=1$ for $n>1$.

Considering a stable particle theory helps better understanding eq.(\ref{factorization}) and (\ref{defZ}). If we take $\phi$ to be a stable particle and $\lambda$ to be an asymptotic one particle state, $\rho_{\lambda}=Z \delta(p^{2}-m^{2})$. Plugging this into eq.(\ref{factorization}), a result from LSZ is recovered. With the same kinds of interactions we are discussing in this letter, $Z$ of the stable particle also vanishes as the (well defined) physical mass approaches one of threshold masses.
On the contrary, from eq.(\ref{factorization}), we see that the production of the particles does not change if off-shell decay modes are also considered. For the stable particles, what is suppressed is the production of on-shell state which remains even after infinite time.

It is natural to define a branching ratio to a channel $\lambda$ as
\bea
{\rm Br_{\lambda}} \equiv \f{\int_{S_{\rm min}}^{S_{\rm max}} dS \, \rho_{\lambda}(S)}{\sum_{\alpha \in \Lambda}\int_{S_{\rm min}}^{S_{\rm max}} dS \, \rho_{\alpha}(S)}.
\label{BR}
\eea 
When we consider only the decay channels whose threshold masses are much below the resonance peak as detectable decay modes, it is easy to see eq.(\ref{BR}) agrees with the conventional definition of branching ratio.

In total, our NWA is
\bea
\sigma(&{\rm ini}\to1,2,\cdots,n, \lambda) \simeq Z' \sigma({\rm ini} \to m_1,\cdots,m_n,m_{\lambda}) \, {\rm Br_{\lambda}}. \nn \\
\label{ourNWA}
\eea
For the decay channels with threshold masses much lower than the physical mass of the unstable particle, $m_{\lambda}$'s correspond to the physical mass of the unstable particle.

\begin{figure*}[t]
\includegraphics[width=0.32\textwidth]{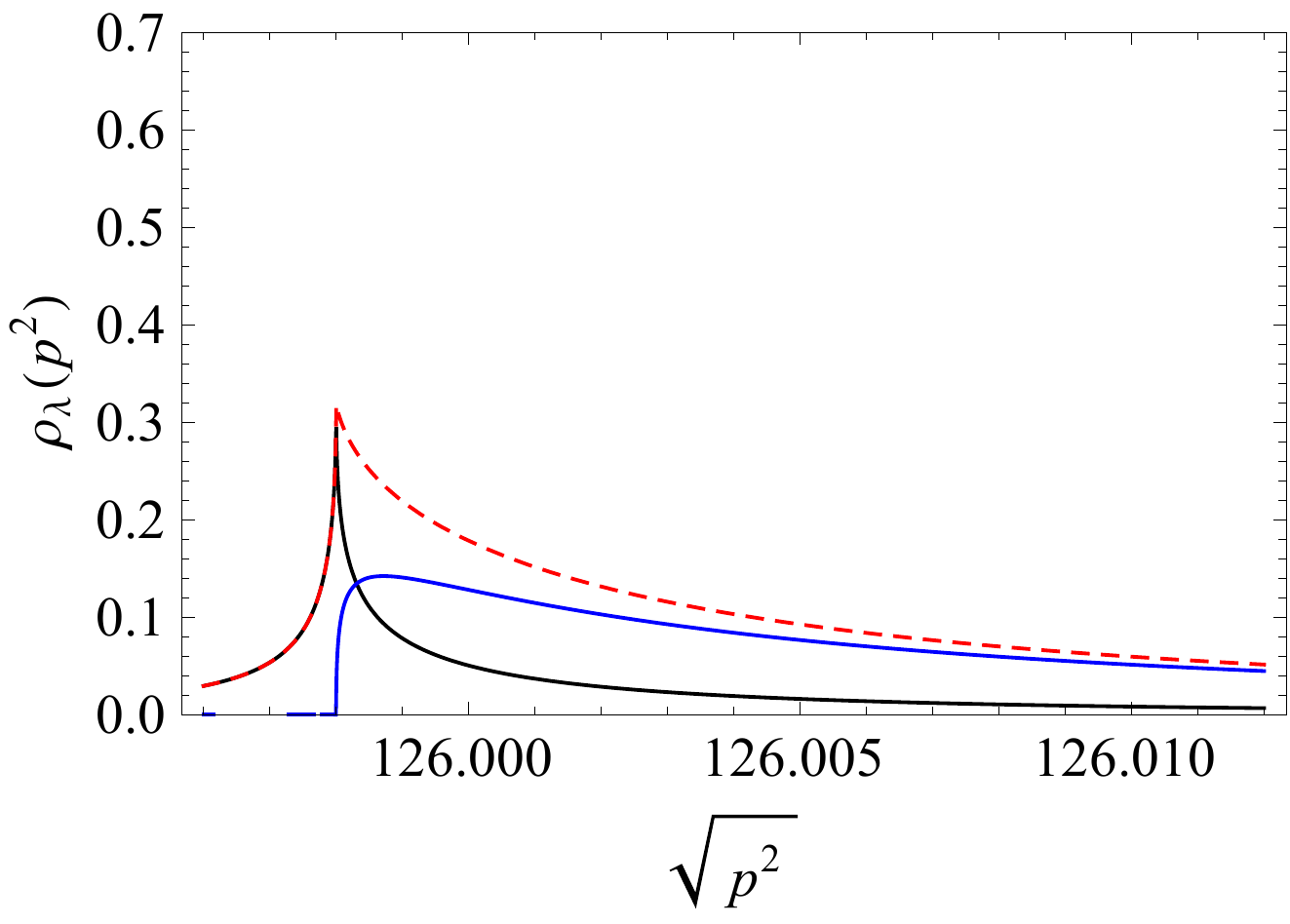}
\includegraphics[width=0.32\textwidth]{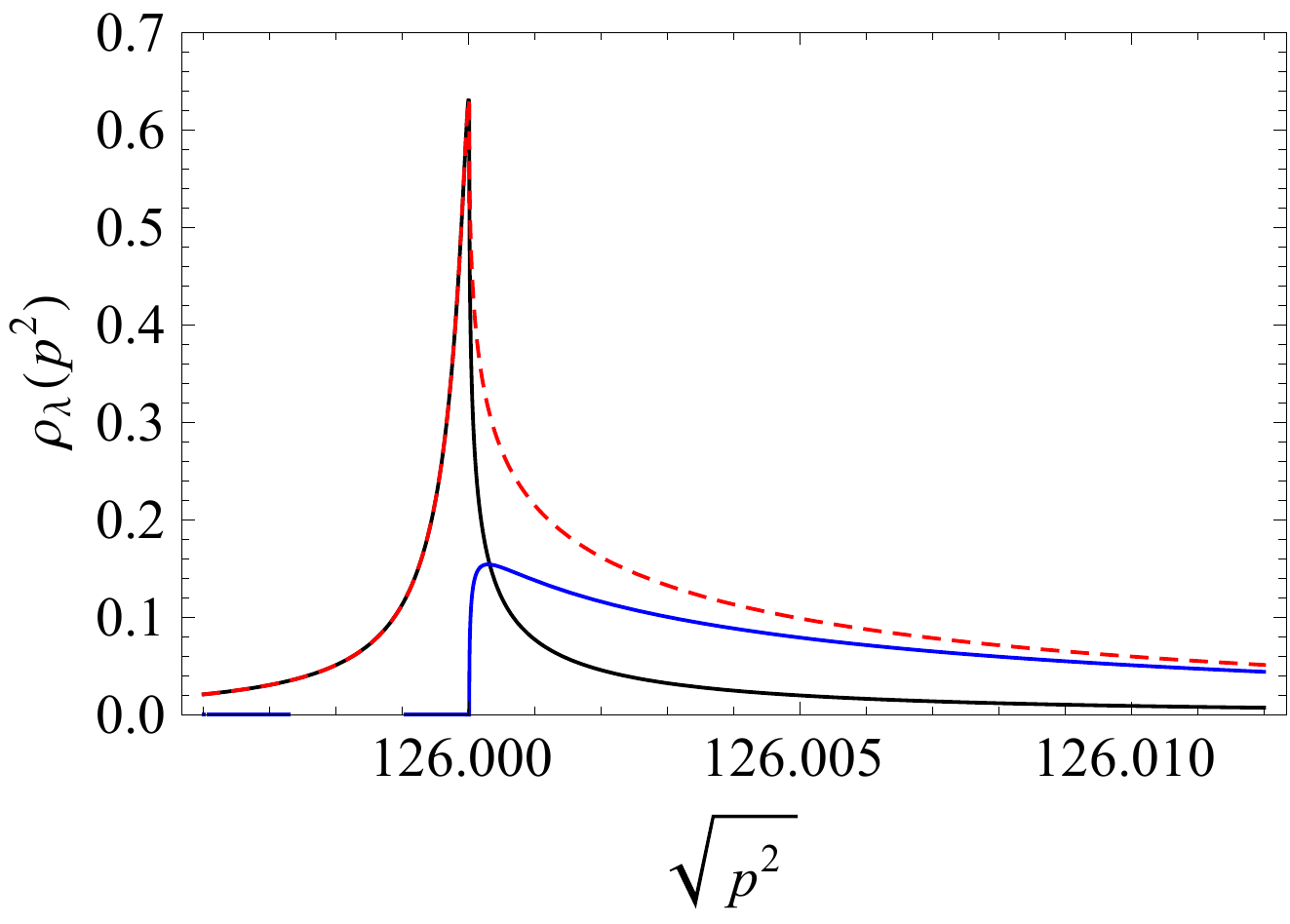}
\includegraphics[width=0.32\textwidth]{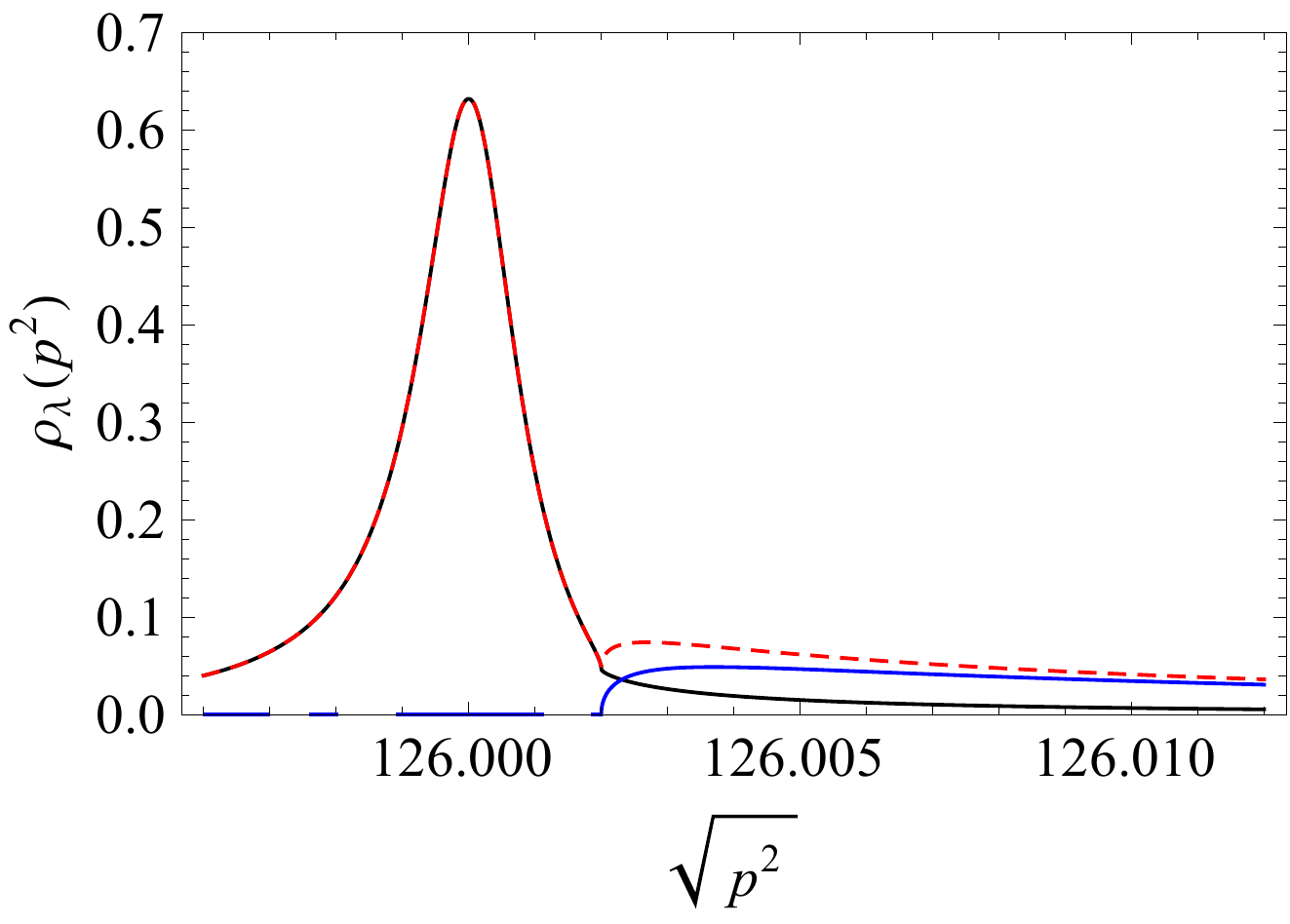}
\caption{
Signal cross sections for each channels are described by spectral density $\rho_\l$. Black curves represent $\rho_{bb}(p^2)$. As the threshold goes to the physical mass, the shape becomes narrower and narrower. The peak position is near the physical mass and its width can be approximated by $\Gamma_1$ in eq.(\ref{g1}). Blue(grey) curves show $\rho_{ss}(p^2)$ and it is zero at and below the threshold. Its width is as large as $\Gamma_2$ in eq.(\ref{g2}) and the peak position is near the threshold. The parameters are fixed by $m_{\rm ph}=126$ GeV, $A=0.6m_{\rm ph}$, $\Gamma_b=4$ MeV for all the three plots. These numbers are used just for an example and corresponding $\epsilon$ in eq.(\ref{strong}) is 0.155. This parameter set is also used in Fig. \ref{ZZZ} and Fig. \ref{survival}. $2m_s-m_{\rm ph}=-\Gamma_b$, $0$, $\Gamma_b$ are used from the left.
}
\label{shapeofresonance}
\end{figure*}

\noindent
{\bf Scalar trilinear as an example}
We take scalar trilinear interactions as a simple example motivated by the setup explaining electroweak symmetry breaking from radiative corrections given in \cite{Dermisek:2013pta}. The real scalar field $h$ that can decay to a light real scalar $b$ has a scalar trilinear interaction with another real scalar $s$,
\begin{equation}
\Delta{\cal L}=-A h(x) s(x)^2.
\end{equation}
where $A$ is a dimensionful coupling and $h$ and $s$ have physical mass of $m_{\rm ph}$ and renormalized mass of $m_s$, respectively.
The physical mass is defined as a zero of ${\rm Re}(G^{-1})$ by absorbing some parts of the self energy at the given scale.
The remaining quantum correction to the inverse propagator is given by 
\begin{equation}
\frac{|2A|^2}{16\pi^2}\Big(f(p^2,m_s^2)-\text{Re}\big(f(m_{\rm ph}^2,m_s^2)\big)\Big)+i m_{\rm ph} \Gamma_b,
\label{selfenergy}
\end{equation}
where
\bea
f(p^2,&m^2)&=-\frac{1}{2}\int_0^1 dx \log\Big(\frac{x^2 p^2 - x p^2 + m^2 -i \epsilon}{m^2}\Big) \label{Imselfenergy} \\
&=&\left\{ \begin{array}{ll}
1-\frac{1}{2}|\bar{\beta}|\ln\frac{1+|\bar{\beta}|}{1-|\bar{\beta}|}&\textrm{if $p^2<0$}\\
1-|\bar{\beta}|\tan^{-1}\frac{1}{|\bar{\beta}|} & \textrm{if $0<p^2<4m^2$}\\
1-\frac{1}{2}|\bar{\beta}|\big(\ln\frac{1+|\bar{\beta}|}{1-|\bar{\beta}|}-i\pi\big) & \textrm{if $p^2>4m^2$}
\end{array} \right., \nn
\eea
and we approximate the imaginary part from the interaction between $h$ and $b$ as $\Gamma_b$ term in eq.(\ref{selfenergy}).

In order to illustrate the influence of the threshold near the resonance, we focus on the limit of threshold at the resonance, $m_{\rm ph}=2m_s$, which is shown in the second plot of Fig. \ref{shapeofresonance}. Though this example deals with the decay into two identical particles, most of physics discussed here is not affected when the decay products are two different particles as long as their sum is near the resonance.

As seen in Fig. \ref{shapeofresonance}, the spectral density shows asymmetric cusp behavior. This distribution was discussed in \cite{Flatte:1976xu} to understand the $\pi\eta$ system. This effect is visible enough when the interaction of the resonance with the threshold particle is stronger than the remaining ones at the resonance.
More specifically, the following condition is assumed for discussions,
\bea
\epsilon = \frac{m_{\rm ph} \Gamma_b}{\Gamma_s^2} & < 1,
\label{strong}
\eea
where $\Gamma_s = |A|^2/(8\pi m_{\rm ph})$ is the asymptotic partial decay width without phase space suppression.
For $\epsilon \ll 1$, there appears two interesting parameters $\Gamma_1$ and $\Gamma_2$.
\bea
\Gamma_{\rm 1} & = & \epsilon \Gamma_b, \label{g1}\\
\Gamma_{\rm 2} & = & \frac{1}{\epsilon} \Gamma_b. \label{g2}
\eea
Unlike the usual interpretation of decay width coming from the imaginary part of the complex pole 
\cite{Bhattacharya:1991gr},
these correspond to {\it real} deviations of two complex poles of the propagator in the second Riemann sheet from $m_{\rm ph}$.

Since the spectral density is $-\f{1}{\pi}{\rm Im} (G)$, its behavior can be understood by looking at which term in the inverse of dressed propagator dominates as we vary energy just above the resonance where
\begin{equation}
G^{-1}\simeq m_{\rm ph}^2|\bar{\beta}|^2+i m_{\rm ph} \Gamma_s \bar{\beta}+im_{\rm ph} \Gamma_b.
\end{equation}
: $i m_{\rm ph}\Gamma_b$ in the nearest of the resonance($m_{\rm ph}$,$m_{\rm ph}+\Gamma_1$), $i m_{\rm ph} \Gamma_s \bar{\beta}$ near the resonance($m_{\rm ph}+\Gamma_1$, $m_{\rm ph}+\Gamma_2$), and $p^2-m_{\rm ph}^2$ away from the resonance($m_{\rm ph}+\Gamma_2$,$\infty$).
Thus the half-width of $bb$ resonance can be represented by $\Gamma_1$. And the cross section to $ss$ starts to appear from the threshold which is at the resonance and lasts up to $\Gamma_2$ which corresponds to the half-width of ss resonance.

The first (third) plot of Fig. \ref{shapeofresonance} shows the spectral densities with the threshold mass below (above) the physical mass. In the first plot, the cusp point is almost the threshold mass where the imaginary part is minimized. In the third plot, the spectral density to $bb$ is quite similar to BW distribution except that it is narrower than the case with $\Gamma_s=0$. For $m_{\rm ph}-2m_s \gsim \Gamma_2$, the maximum of bb channel resonance is mainly determined by the real part of the self energy and the BW distribution is recovered for bb channels. 

Fig. \ref{ZZZ} describes $Z'(\infty;bb)$, $Z'(\infty;bb,ss)$, and conventional $Z$ as a function of the threshold mass. Although $E_{\rm tot}$ does not have to be infinity, we use it for notational convenience to mean that  the interval ($S_{\rm min}$,  $S_{\rm max}$) sufficiently covers the resonance region. As the threshold mass approaches the physical mass from above, the dotted line in Fig. \ref{ZZZ}, Z goes to zero. Thus, the conventional NWA shows a pathological behavior of the scalar $h$ at the threshold, which is vanishing full scattering cross section involving $bb$ channel. In contrast, the definition in eq.(\ref{defZ}) is well defined even in this limit. For $bb$ channel, $Z'(\infty;bb)\simeq \epsilon$ is small but non-zero at the physical mass and it gives correct approximation to the full scattering cross section combined with eq.(\ref{ourNWA}). Alternatively, we can start from $Z'(\infty;bb,ss)=1$ now with ${\rm Br_{bb}}\simeq \epsilon$ by including ss channel. Both scenarios explain that the full cross section to bb channel is suppressed due to the threshold near the resonance.

It is interesting to ask which one of widths is the decay rate among $\Gamma_b$, $\Gamma_s$, $\Gamma_1$, and $\Gamma_2$. A survival probability is actually a concept depending on how unstable particles are prepared, or mass filtered \cite{Schwinger:1960}. Here, we consider an ideal situation in which produced unstable particles sufficiently cover the resonance region so that we can use $P(t)\equiv \left|\int_0^\infty dS e^{-i \sqrt{S} t} \rho (S) \right|^2$.

%Around the resonance, the Fourier transform of the spectral density gives factor $e^{-i m_{\rm ph} t}$ which is an overall phase and of no interest. Therefore, the resonance part gives very slowly varying piece in the survival amplitude and probability
Back to the limit $2m_s=m_{\rm ph}$,
 major decay pattern can be understood by saperating the  energy region ($m_{\rm ph}+\Gamma_1$, $m_{\rm ph}+\Gamma_2$) and ($m_{\rm ph}+\Gamma_2$, $2m_{\rm ph}$) where the spectral density $2E\rho(E^2)$ can be approximated by $(\f{1}{\pi} \sqrt{\f{2}{\Gamma_2}})\f{1}{\sqrt{E-m_{\rm ph}}}$ and $(\f{1}{\pi} \sqrt{\f{\Gamma_2}{2}})(\f{1}{\sqrt{E-m_{\rm ph}}})^{3}$, respectively. Roughly speaking, their effects on the survival probability P(t) are such that two characteristic patterns of non-exponential decay appear, $1/(\Gamma_2 t)$ at time much later than $1/\Gamma_2$, and $e^{-\sqrt{\Gamma_2 t}}$ at time before $1/\Gamma_2$.

Numerically exact result and analytic approximation of P(t) are shown in Fig. \ref{survival}. The approximation is done by summing up those two functions approximated in the above paragraph with exponentially suppressing envelop. The plot shows that the decay is non-exponential. To help make a contrast with exponential decay, staright dotted lines are introduced in the plot. The reason why the plot looks straight at large $t$ is that exponential and power functions are difficult to distinguish for short interval at large $t$ in log plot.

\begin{figure}[t]
\includegraphics[width=0.4\textwidth]{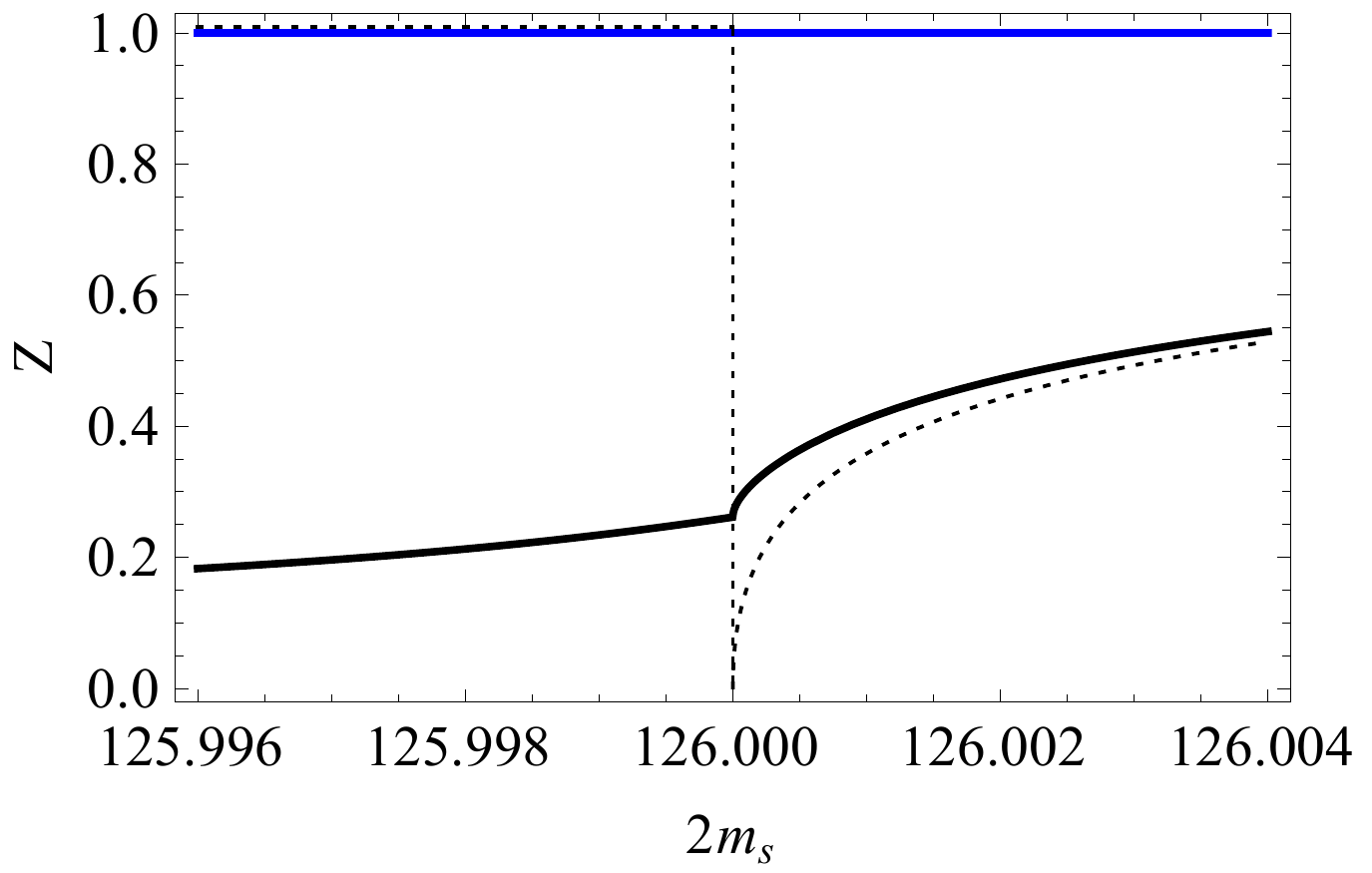}
\caption{ $Z'(\infty;bb)$ (black), $Z'(\infty;bb,ss)$ (blue/grey), and conventional $Z$ (Dotted) as a function of the threshold mass. The conventional $Z$ is ill-defined as its left limit and the right limit do not match at $2m_s=m_{\rm ph}$ while $Z'$s are well-defined in all range. $Z'(\infty;bb)$ asymptotically reaches to $Z$ as $2m_s$ goes far above the resonance.}
\label{ZZZ}
\end{figure}

\noindent
{\bf Classification}
In this section, we list possible types of interactions which can cause conventional $Z$ ill-defined such that the discussion given in this Letter is relevant. The key point is from eq.(\ref{Imselfenergy}) that the leading order expansion of self energy around $\sqrt{p^2}=\sum_i m_i$ is $\hat{\beta}\equiv\sqrt{1-\frac{(\sum_i m_i)}{\sqrt{p^2}}}$. Thus if $m_{\rm ph}$ is accidentally around $\sum_i m_i$, then one loop correction overcomes the tree level part, $p^2-m_{\rm ph}^2\simeq 2m_{\rm ph}^2 \hat{\beta}^2$. The leading power of $\hat{\beta}$ in the expansion of one loop correction determines the behavior. If the leading power is lower than one, then Z vanishes as physical mass and threshold get closer and the ill-defined $Z$ should be cured with the method suggested in this Letter.

First, in scalar trilinear interaction, even if all the three scalar fields are different species, the leading order expansion is linear in $\hat{\beta}$.
For Yukawa interaction, $S \bar{\psi}_1 \psi_2$ among the scalar $S$ and fermions $\psi_i$, only the self energy of fermion field has a term linear in $\hat{\beta}$, while the self energy of scalar has $\hat{\beta}^3$ due to extra factor from spinor.
If the Yukawa interaction involves $\gamma_5$, such that $S \bar{\psi}_1 \gamma_5 \psi_2$, then only the self energy of (pseudo-)scalar field has the problematic expansion in which the leading power is square root order.
%Details of NWA for polarization/spinor are described in \cite{Uhlemann:2008pm}.

All those three cases significantly change resonance shape from BW form as we discussed in this letter. We leave gauge interaction cases as a future work.

\begin{figure}[t]
\includegraphics[width=0.4\textwidth]{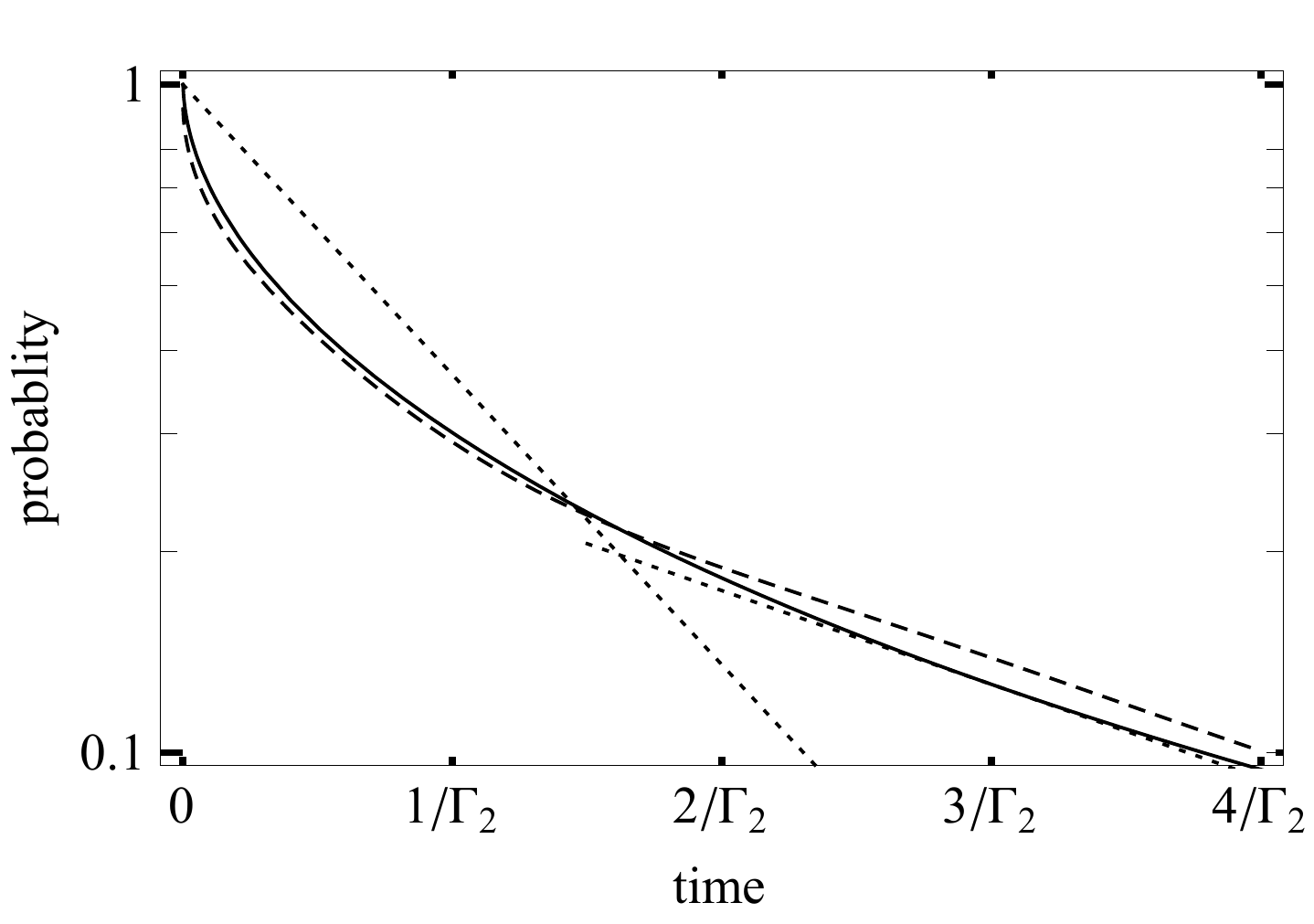}
\caption{Survival probability, P(t) obtained by a numerical way (Solid line), analytical approximation (dashed) and exponential decay for comparison(dotted).}
\label{survival}
\end{figure}

\noindent
{\bf Conclusion}
We discussed physics of particle production and decay when the threshold is at or near the resonance. The conventional wave function renormalization factor or the derivative of the self energy is ill-defined, and usual Breit-Wigner approximation fails since the self energy varies rapidly right at the resonance. Different decay modes show different resonance shapes including the narrowly peaked resonance and the broad threshold resonance. Apparently there is no way to use the narrow width approximation due to complicated physics at around the resonance in this case. 

In this Letter, we showed that we can still define narrow width approximation given in eq.(\ref{ourNWA}) if (i) the production cross section does not change rapidly within the the resonance, (ii) the resonance width is small enough compared to the produced invariant mass $S$ range, and (iii) the branching ratio is defined as the integral of the specific channel spectral density as in eq.(\ref{BR}). There is no unique decay width defining the unstable particle and it should be found from the shape of the resonance (half-width) or from the decay pattern (inverse of half-lifetime). The width is then channel dependent. The same is true for the resonance peak position or the mass of the unstable particle.

The effort to look for unanimous definition of mass and width for unstable particles would be in vain as the study in this Letter shows. Only stable particles can appear as asymptotic states. Whenever pathological problems appear, the full self energy of the unstable particles should be used instead of BW approximation. Still the generalized NWA works, and the threshold at the resonance reduces the branching ratio to the other decay channels below the physical mass. The influence of new states at the resonance or slightly above is correctly taken account in the generalized NWA.

\noindent
{\bf Acknowledgements}
This work was supported by the National Research Foundation of Korea(NRF) No. 0426-20140009.
HK thanks  Paul Langacker, Juan Maldacena and Edward Witten for discussions.
HK is supported by IBM Einstein fellowship of Institute for Advanced Study. DC and TJ thank Institute for Advanced Study for its warm hospitality during the visit.

\end{document}